\documentclass[%
 aip,
 amsmath,amssymb,
 reprint,%
]{revtex4-1}

\usepackage{graphicx}
\usepackage{dcolumn}
\usepackage{bm}
\usepackage[utf8]{inputenc} 
\usepackage[T1]{fontenc}
\usepackage{mathptmx}

\begin{document}

\preprint{AIP/123-QED}

\title{Prospects and challenges of quantum emitters in 2D materials}

\author{Shaimaa I. Azzam}
\altaffiliation{These authors contributed equally to this work.}
\affiliation{Department of Electrical and Computer Engineering, University of California, Santa Barbara, CA 93106 }
\affiliation{California Nanosystems Institute, University of California Santa Barbara, CA 93106}

\author{Kamyar Parto}%
\altaffiliation{These authors contributed equally to this work.}
\affiliation{Department of Electrical and Computer Engineering, University of California, Santa Barbara, CA 93106 }

\author{Galan Moody}
\email{moody@ucsb.edu}
\affiliation{Department of Electrical and Computer Engineering, University of California, Santa Barbara, CA 93106 }
\affiliation{California Nanosystems Institute, University of California Santa Barbara, CA 93106}

\date{\today}

\begin{abstract}
The search for an ideal single-photon source has generated significant interest in discovering novel emitters in materials as well as developing new manipulation techniques to gain better control over the emitters' properties.  Quantum emitters in atomically thin two-dimensional (2D) materials have proven very attractive with high brightness, operation under ambient conditions, and the ability to be integrated with a wide range of electronic and photonic platforms. This perspective highlights some of the recent advances in quantum light generation from 2D materials, focusing on hexagonal boron nitride and transition metal dichalcogenides (TMDs).  Efforts in engineering and deterministically creating arrays of quantum emitters in 2D materials, their electrical excitation, and their integration with photonic devices are discussed. Lastly, we address some of the challenges the field is facing and the near-term efforts to tackle them. We provide an outlook towards efficient and scalable quantum light generation from 2D materials towards controllable and addressable on-chip quantum sources.
\end{abstract}

\maketitle
\section{\label{sec:intro}Introduction}
  
Two-dimensional (2D) materials, such as graphene, hexagonal boron nitride, and transition metal dichalcogenides, are a nascent family of materials. 2D materials exhibit quantum properties that are generally absent in their bulk counterparts; this includes a layer-dependent bandgap \cite{chaves2020bandgap}, large exciton binding energies \cite{manzeli20172d},  strong nonlinearities,  tunable valley degree of freedom, the ability to host quantum emitters\cite{toth2019single} and spin-defects\cite{turiansky2020spinning} 
Moreover, due to their atomic thickness, 2D materials can readily be integrated with electronic and photonic devices, facilitating precise engineering of light-matter interaction at the nanoscale.   The large library of available 2D materials \cite{manzeli20172d}, combined with the ability to stack them with precisely controlled alignment and orientation \cite{liu2016van}, provide a unique test-bed for atomically smooth and thin heterostructures, known as Van der Waals (vdW) heterostructures, with well-controlled and tunable optoelectronic properties and quantum confinement.   
This unique set of properties have turned the 2D materials into an exciting  test-bed for exploring novel quantum phenomena  such as  quantum light generation emitters\cite{toth2019single}, spin-qubit applications, valley-spintronics\cite{schaibley2016valleytronics}, and twisted moire superlattices for engineering correlated many-body physics\cite{manzeli20172d,cao2018unconventional,serlin2020intrinsic,tran2019evidence}.

\begin{figure*}[t] \centering
     \includegraphics[scale= 1.5]{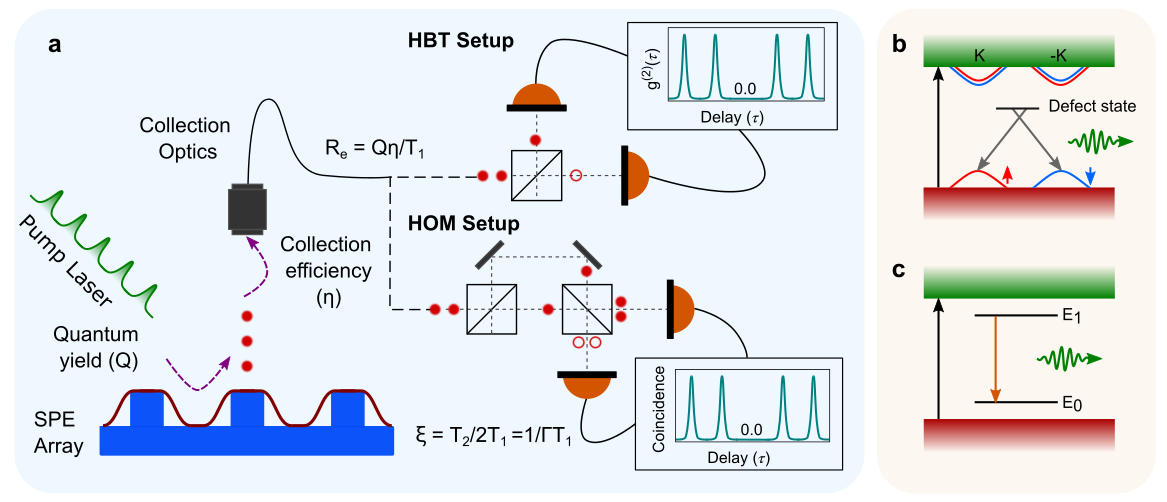}  
          \caption{Single-photon emission and characterization in 2D materials. a) An illustration of the characterization of the  single photons generated from an array of SPEs. The Hanbury-Brown and Twiss (HBT) setup is used to measure the purity ($g^2(0)$) of the generated photons whilst a Hong-Ou-Mandel (HOM) interferometer is used to determine their indistinguishability ($\xi$).
          b) Schematic  of single-photon generation from   localized excitons in TMDs (e.g. WSe$_2$). The strain localized excitons hybridize with point defects leading to valley selectivity breaking and strong photoemission.  c) Single-photon generation from deep defect levels ($E_0$ and $E_1$) in wide bandgap insulator (e.g. hBN). }  
          \label{fig:intro}
\end{figure*}

Engineering quantum confinement in 2D materials has attracted particular interest in recent years, with several seminal papers demonstrating atomic defect-based single-photon emitters (SPEs) in transition metal dichalcogenides and hexagonal boron nitride. SPEs are at the heart of numerous quantum technologies, including quantum cryptography,  quantum communication, quantum information, and quantum sensing. An ideal SPE must meet several metrics set by the requirement of their targeted applications. The most critical metrics can be summarized as (1) brightness ($R_e)$) that determines the rate of photons that can be extracted from the system, on-demand, which can be expressed as a product of the rate of the incident pulsed laser multiplied by quantum yield ($Q$) and collection efficiency ($\eta$). However, note that the radiative lifetime ($T_1$) of the emitter sets the maximum extractable rate of the system \ref{fig:intro}(a)  ; (2) single-photon purity, defined by the value of second-order autocorrelation function at zero time delay $g^2(0)$, which quantifies the "one at a time" behavior and sub-Poissonian, non-classical nature of the SPE. The purity  is generally measured through a Hanbury-Brown and Twiss (HBT) interferometer as shown in Fig. \ref{fig:intro}(a).  Importantly, a simultaneously high brightness and purity are required to avoid a vacuum component with high probability amplitude; (3) indistinguishability ($\xi$), determined through Hong-Ou-Mandel (HOM) interferometry, depicted in Fig. \ref{fig:intro}(a),  quantifies how well-defined the spatio-temporal mode is of two photons emitted from the same or different emitters through measurement of the two-photon interference visibility; and (4) stability and reproducibility, which includes working temperatures, blinking, bleaching, spectral wandering, site control, etc., sets a practical limit in terms of yield, scalability, and reliability of each technology. Most quantum photonic applications require purities ($g^2(0)$) well below 0.01, indistinguishability ($\xi$) above 0.99, extraction efficiencies ($Q.\eta$) above 0.99, and brightness in gigahertz \cite{aharonovich2016solid}. To this date, no SPE technology has met all the metrics required for scalable quantum information applications with SPEs \cite{aharonovich2016solid}.

The search for an ideal SPE has fueled considerable interest spanning multiple platforms, with quantum dots and defect centers in solids being the most investigated candidates. However, despite the rapid progress in recent years, these material platforms face several challenges, some of which are intrinsic to their host materials, motivating the search for SPEs in alternative platforms. For instance, self-assembled InAs quantum dots \cite{somaschi2016near,ding2016demand} are considered state-of-the-art in terms of purity and indistinguishability, yet spatial and spectral inhomogeneity has prevented the development of large scale arrays of identical emitters. On the other hand, defect-based emitters in wide bandgap materials offer a more direct route towards site-controlled placement using defect engineering techniques, but they are more challenging to integrate with available photonic technologies, tend to have low photon extraction efficiencies, and are challenging to address electrically.

Recently, single-photon emission has been reported from semiconducting 2D , namely transition metal dichalcogenides (TMDs) \cite{srivastava2015optically, he2015single, chakraborty2015voltage, koperski2015single, tonndorf2015single}, insulating 2D hexagonal boron nitride (hBN) \cite{tran2016quantum}, thin-film gallium selenide \cite{tonndorf2017chip}, and  moir{\'e}-trapped excitons \cite{baek2020highly}, promising unique advantages compared to other solid-state emitters. Due to their atomic thickness and dangling bond-free interfaces, quantum light extraction from 2D materials is highly efficient and the modulation of their properties is more straightforward through vdW heterostructure stacking. Emitters in atomically thin host materials can be more easily accessed and interfaced to judiciously designed electronic and photonic devices to facilitate their integration. Moreover, as recent experiments have shown, emitters in 2D materials are well-suited for realizing deterministic arrays of SPEs that can be externally addressed with electrically controllable switching of emission from the SPEs \cite{hotgergate2021gate}.   

Within the last few years, the quality, yield, tunability, and physical origins of SPEs in 2D materials have rapidly advanced. This perspective focuses on shining light on these advances within the lens of scalability and technological relevance. We provide an overview of key methods for the deterministic fabrication of SPEs through defect and strain engineering, and we discuss unique approaches to external tuning and addressability. We conclude with the challenges the field is facing and the future opportunities towards achieving scalable arrays of identical SPEs, providing a concise roadmap for where this field may make the largest impact in quantum information science. For more in-depth reviews on the physics of SPEs in 2D materials and other platforms, we refer the readers to reviews found in \cite{aharonovich2016solid, toth2019single, lee2020integrated}.


\section{Single-photon emitters in 2D materials}

\begin{figure*}[t] \centering
     \includegraphics[scale= 0.5]{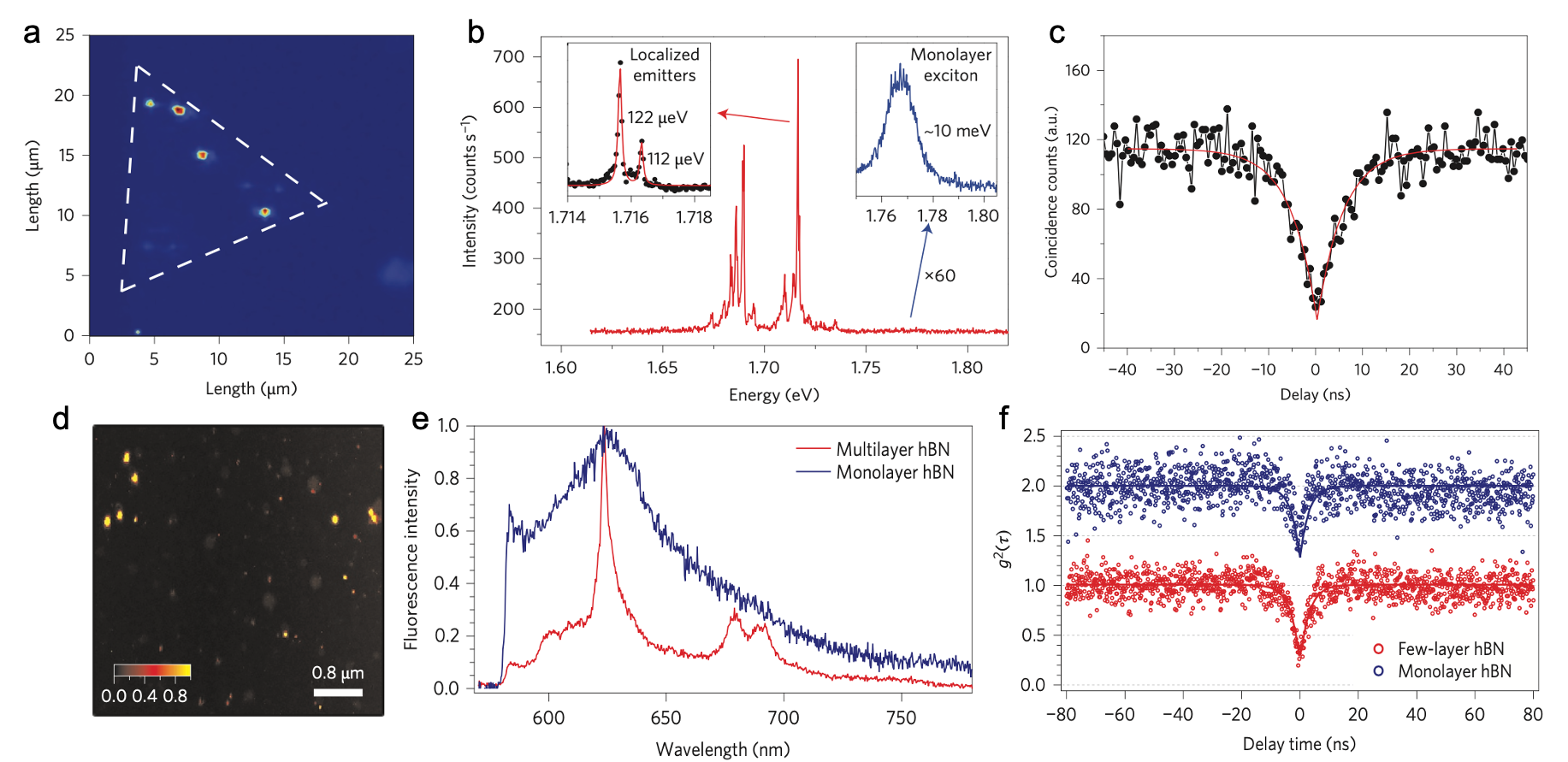}  
          \caption{Quantum emitters in 2D materials. a)  A map of PL intensity showing the  narrow emission lines of the emitters in monolayer WSe$_2$ \cite{he2015single}. The dashed triangle shows the WSe$_2$ monolayer. b) PL spectrum of the quantum emitters. The left inset shows the spectrum of the highest intensity peak and the right inset indicated the PL from the amplified monolayer valley exciton emission. c) Second-order correlation measurement of the emitter in (b) showing   \( g^2(0) \) = 0.14 $\pm$ 0.04.  d)  Scanning confocal map of an hBN sample showing bright spots, some of which are quantum emitters \cite{tran2016quantum}. e) Room-temperature PL of quantum emitters in hBN monolayer (blue) and multilayer (red). f)  Second-order correlation measurement of the emitters in (e) confirming single-photon emission from monolayer (blue) and multilayer (red) hBN samples. 
          Panels (a-c): Reprinted by permission from \cite{he2015single}. Copyright 2015, Springer Nature. Panels (d-f): Reprinted by permission from \cite{tran2016quantum}. Copyright 2015, Springer Nature. 
          }   
          \label{fig:SPEs_2D}
\end{figure*}

Single-photon emitters (SPEs) have been identified in several types of 2D materials including semiconducting \cite{srivastava2015optically, he2015single, chakraborty2015voltage, koperski2015single, tonndorf2015single}, such as tungsten diselenide (WSe$_2$), tungsten disulfide (WS$_2$), and molybdenum diselenide (WSe$_2$), as well as insulating hBN \cite{tran2016quantum, proscia2018near, spokoyny2020effect, konthasinghe2019rabi, jungwirth2016temperature}. 
However, the microscopic origin of the quantum emitter is different depending on the host material. The origins of the SPEs in TMDs are still under extensive experimental and theoretical investigations. One current hypothesis is that the quantum emission in monolayer TMDs originates from radiative recombination of the WSe$_2$ dark exciton ground-state through an intermediate localized state in regions of high strain. \cite{linhart2019localized}. These localized states have been attributed to crystal imperfections of crystallographic defects as illustrated by the schematic in Fig. \ref{fig:intro}(b)\cite{zheng2019point, parto2020irradiation}. This model captures all the essential features of SPEs first observed in monolayer WSe$_2$ in 2015 \cite{srivastava2015optically, he2015single, chakraborty2015voltage, koperski2015single, tonndorf2015single}. In these studies, SPEs appear as sharp lines in the photoluminescence (PL) spectrum at cryogenic temperatures. Generally, the emitters appear at random locations in the TMD monolayer with a strong correlation to appear at the edges or in proximity to wrinkles on the flake. Examples of quantum emitters in  WSe$_2$ are shown in  Fig. \ref{fig:SPEs_2D}(a) where the bright localized spots correspond to the SPEs; however, within the last few years, the random emitter sites has been controlled to a degree using numerous techniques for the deterministic creation of defects via, for example, edge creation \cite{ziegler2019deterministic}, or local strain \cite{palacios2017large, branny2017deterministic, parto2020irradiation}, some of which has succeeded at achieving close to unity yield. SPEs with qualitatively similar features have been observed in other TMDs including WS$_2$ and MoSe$_2$.

The low-temperature photoluminescence (PL) of SPEs is spectrally narrow (FWHM 100 $\mu eV$) compared to the of unbound excitons as can be seen in Fig. \ref{fig:SPEs_2D}(b) with lifetimes of about 2 ns and they tend to be red-shifted with respect to the free exciton. The fact that the $\sim100$ meV linewidth is not lifetime-limited suggests that significant spectral diffusion or wandering is present; however, at present, the intrinsic homogeneous linewidth (dephasing rate) has not been measured for TMD-based SPEs. Nonetheless, the emitters' quantum nature has been verified using second-order correlation function under continuous-wave excitation, as shown in Fig. \ref{fig:SPEs_2D}(c), and also using pulsed-wave excitation, proving that a single photon can be generated on-demand from WSe$_2$ monolayers. Thus far, the majority of studies on SPEs from TMDs have been limited to cryogenic temperatures; however, recently, the operating temperature of reported emitters has significantly improved with singe photon emission sustained up to 160 K  \cite{luo2019single, parto2020irradiation} even without Purcell enhancement, suggesting that room-temperature operation may be feasible with further material engineering and photonic or plasmonic integration.

Following the discovery of SPEs in semiconducting 2D materials, single-photon emission from hBN monolayers and multilayers has been reported at room temperature \cite{tran2016quantum}, as illustrated in Fig. \ref{fig:SPEs_2D}(d-f). The microscopic origins of SPEs in hBN are similar to optical defect transitions observed in other wide-gap insulators, such as diamond and silicon carbide defect centers. hBN has a large bandgap of around 6 eV, which hosts a large range of optically active defects spanning near-infrared to ultraviolet energies (Fig. \ref{fig:intro}(c)).  The Debye-Waller (DW) factor, which expresses the ratio of emission into the zero-phonon line (ZPL) to that of the total integrated emission intensity, is significant for hBN, reaching up to 82\%. DW factor values for hBN are comparable to that of the negative silicon-vacancy ($Si_V^{-1}$) center in diamond \cite{dietrich2014isotopically} and much higher than that of the negatively charged nitrogen-vacancy ($N_V^{-1}$) center in diamond, with a DW factor of only 4\%, showing a promise of  hBN a significant contender as a host for SPEs.
Quantum emitters in monolayer hBN are spectrally broadened beyond the radiative limit at room temperature (blue curve in Fig. \ref{fig:SPEs_2D}(e)), which negatively impacts their indistinguishability. They also suffer from blinking and bleaching over time \cite{tran2016quantum}. However, multilayer hBN samples show a more robust behavior with no blinking or bleaching and a sharper spectral response \cite{tran2016quantum}. 

One of the significant challenges with hBN SPEs is the considerable variability in the ZPL wavelengths and their spectral shapes. As mentioned earlier, hBN emits in a wide range of energies spanning near bandgap emission in the 4-6 eV ultraviolet regime, and defect levels deep in the bandgap emitting in the visible regime around 2 eV with a considerable spread. This spectral heterogeneity represents a major challenge in the production of scalable and robust SPE and a serious hurdle facing their integration with electronic and photonic devices. In addition, due to the wide range of reported ZPL energies and properties, the origins of the emitters in hBN are a topic of much debate. Numerous studies have examined plausible mechanisms responsible for hBN emitters, including vacancy point defects and impurities. A neutral nitrogen atom occupying a missing boron site adjacent to a nitrogen-vacancy ($V_N N_B $) and the negatively charged boron vacancy ($V_B^-$) were a candidate for the visible single-photon emission \cite{tran2016quantum}. Similarly, a neutral and positively-charged nitrogen-vacancy adjacent to a carbon impurity on a boron site ($V_N N_B$) \cite{abdi2018color}, and a boron vacancy adjacent to two oxygen atoms ($V_B O_2$) \cite{xu2018single}, have also been suggested as the possible origin of the quantum emitters. Other reports suggest the dangling bond in hBN to be the reason for the SPEs in the visible due to a doubly occupied boron dangling bond \cite{turiansky2019dangling}. More recently, experimental evidence suggests that carbon impurities are the source of visible SPEs through the negatively charged $V_B C_N$ \cite{mendelson2020identifying}. 

\section{\label{sec:research}Recent advances in defect engineering}
Since the discovery of quantum emitters in 2D materials, tremendous efforts have been directed to better understand their origins as well as to engineer their creation and properties. This section highlights some of the recent advances in the engineering and control of SPEs in 2D materials. Specifically, it focuses on the efforts on the deterministic fabrication of SPEs through strain engineering and accelerated-beam irradiation, their electrical control and addressability, and integration with photonic devices. 

\subsection{Engineering of SPEs}

\begin{figure*}[t!] \centering
     \includegraphics[scale= 0.6]{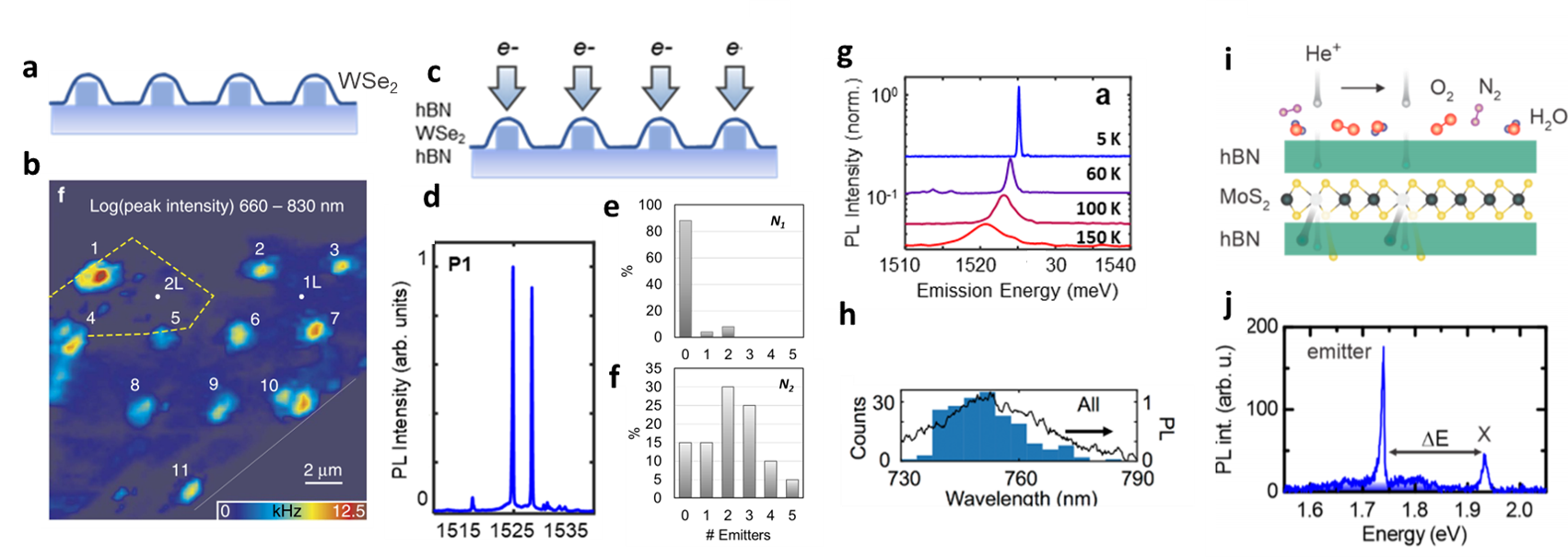}  
          \caption{Advances in the engineering of quantum emitters. a) Illustration of WSe$_2$ layer on an array of nano-pillars. b) PL map of a stamped WSe$_2$ flake on a nan-opillar array \cite{branny2017deterministic}. A bright emission from localized emitters on top of nano-pillar is observed. c) Illustration of strain and defect engineering process. The strain is first engineered through nano-pillar stressors; the top of the pillars are then irradiated with electron beam to induce defects. d) PL spectrum of a strained and irradiated site demonstrating sharp localized exciton and biexciton features \cite{parto2020irradiation}. e) Percentage of quantum emitters found on each pillar site when irradiated with e-beam dosage of $10^5$ electron $/ \mu m^2$.  f) Percentage of quantum emitters found on each pillar site when irradiated with e-beam dosage of $10^6$ electron $/ \mu m^2$. An average number of emitters per site increases as a function of e-beam intensity and induced emitters. g) PL spectrum of WSe$_2$ emitters as a function of temperature. WSe$_2$ emitters preserve their \( g^2(0) \)<0.5 up to 160 K \cite{parto2020irradiation}. h) A spectral histograms of engineered SPEs showing a recreattion of the lineshape of the low-temperature broad defect band of WSe$_2$ \cite{rosenberger2019quantum}. i) Engineering SPEs in hBN encapsulated MoS$_2$ using Helium ion beam. hBN reduces the surface adsorbents and improves the spectral linewidth and stability of the emitters \cite{klein2020scalable}. j) PL spectrum of the hBN encapsulated MoS$_2$ emitters. SPEs are detuned approximately 200 meV below the free exciton with purities as high as 0.8.
          Panel (h): Reprinted  with permission from \cite{rosenberger2019quantum}. Copyright 2019, American Chemical Society.
          }
          
          \label{fig:SPE_engineering}
\end{figure*}

Shortly after observing the seemingly randomly distributed SPEs in WSe$_2$, initial attempts were made to understand and control their creation. Kumar \textit{et al}.\cite{Kumar2015strain} and Branny \textit{et al}.\cite{Branny2016MoSe2} proved that SPEs have a high tendency to appear in strained regions of TMDs, and their spectral properties are sensitive to the magnitude of the strain. This was further validated by Kern \textit{et al}.\cite{kern2016nanoscale}, who were successful at semi-deterministic creation of these emitters by laying them across gaps between gold nanorods. In 2016, site-specific engineering of emitters reached maturity by the demonstration of Branny \textit{et al}.\cite{branny2017deterministic} and Palacious \textit{et al}.\cite{palacios2017large}. By stamping TMDs onto a SiO$_2$ substrate decorated with sub-$\mu$m sharp nano-pillars, which acted as nano-stressors, TMDs were strained in a highly localized manner as seen in Fig. \ref{fig:SPE_engineering}(a). This method achieved almost a near-unity yield in the creation of site-specific emitters (Fig.\ref{fig:SPE_engineering}(b)), with purities reaching as high as 95$\%$ and detection rates up to 10 kHz; 
however, a withstanding challenge was that emitters were still appearing in a large spectral range of 720-800 nm, and also, these emitters were limited to cryogenic temperatures. Other strain-engineering methods have been explored following these demonstrations, such as metallic nano-cubes\cite{luo2018deterministic} and nano-particles\cite{peng2020creation}, nano-indentation with AFM tips\cite{rosenberger2019quantum}, and electrically controlled microcantilevers\cite{kim2019position}. While each of these approaches have advantages, such as dynamical control over the extent of strain\cite{kim2019position} or concurrently achieving Purcell enhancement\cite{luo2018deterministic,peng2020creation}, overall, the emitters' purities were not comparable to nano-pillar approaches\cite{branny2017deterministic,palacios2017large}. For each of these approaches to strain engineering, control over the SPE emission energy has not been demonstrated, and SPEs appear across a large distribution of energies.

While strain engineering methods were successful at site-specific engineering of SPEs, however, it was long hypothesized that defects also play a fundamental role in these SPEs' microscopic origins \cite{tonndorf2015single,Branny2016MoSe2}. The theoretical study by Linhart \textit{et al}.\cite{linhart2019localized} suggested that both defects and strain are required to explain the physical properties of SPEs in WSe$_2$ and answer fundamental questions pertaining to why these emitters exhibit bright emission. Further studies corroborated pieces of this picture. For instance, Liu \textit{et al.}\cite{luo2018deterministic,luo2019single} observed that WSe$_2$ monolayers with low defect density increase the working temperatures of their device. Moon \textit{et al}.\cite{moon2020strain}, by careful examination of the spectral range of their emitters with varied strain, observed that no emitters appear below the predicted dark-exciton binding energy of WSe$_2$. Moreover, Branny et al.\cite{branny2017deterministic} and Rosenberg \textit{et al}.\cite{rosenberger2019quantum} observed that the spectral histograms of their engineered SPEs recreate the lineshape of the well-known low-temperature broad defect band of WSe$_2$, suggesting a correlation between defects and emitters (Fig. \ref{fig:SPE_engineering}(h)). In our recent study\cite{parto2020irradiation}, additional evidence supporting this model was observed whereby using electron beam irradiation to induce defects in TMDs whilst using nano-pillars to engineer strain (Fig. \ref{fig:SPE_engineering}(c,d)), a direct correlation between the intensity of the defect band and the number of quantum emitters engineered in WSe$_2$ exists (Fig. \ref{fig:SPE_engineering}(e,f)). This approach essentially decouples the strain and defect engineering steps and allows for better control of the SPE engineering process. For instance, electron beam (e-beam) induced defects appeared at $\sim$100-150 meV lower energies leading to higher thermal energy barriers, which allowed emitters to remain functional up to 150 K without Purcell enhancement as demonstrated in Fig. \ref{fig:SPE_engineering}(g). Furthermore, the emitters' purities in this approach reached to 95\%, and a radiative cascade indicative of exciton-biexciton emission was observed from some emitter sites. More recently, Klein \textit{et al}.\cite{klein2019site,klein2020scalable} were also successful at creating SPEs in MoS$_2$ using helium-ion irradiation to create defects without inducing any strain (Fig. \ref{fig:SPE_engineering}(i,j)). Given that the spin-orbit coupling in MoS$_2$, leading to splitting between dark and bright excitons, is not as large as WSe$_2$, it is up to debate whether these emitters originate solely from defects or whether they form within a similar framework as intervalley defect excitons\cite{linhart2019localized}. Overall, the interplay between defects, excitons, and electronic structures of TMDs gives rise to a rich optoelectronic platform that demands further interrogation.

Similar SPE engineering efforts are also underway in the hBN system, where the defect transitions are responsible for single-photon emission; however, a key challenge here is distinguishing whether the engineering techniques alter an intrinsic defect present in the material to luminescence (defect activation) , or whether a new defect is created within the process. Efforts such as thermal annealing \cite{tran2016robust,choi2016engineering}, chemical etching \cite{chejanovsky2016structural}, ion irradiation \cite{choi2016engineering,chejanovsky2016structural}, plasma treatment \cite{xu2018single,vogl2018fabrication}, e-beam irradiation \cite{tran2016robust,choi2016engineering}, and laser irradiation \cite{choi2016engineering} have been shown to increase the SPE concentration in hBN flakes. Interestingly, nano-pillars were also successful, although with a lower yield than TMDs, in site-specific engineering of emitters in hBN \cite{proscia2018near}. Considering that hBN emitters tend to appear mostly close to the edge of the flake, edge creation methods such as ion-beam milling have improved the yield of site-specific emitter engineering processes \cite{ziegler2019deterministic}. Notably, a recent e-beam irradiation method has been able to achieve a high yield, high spatial accuracy, and spectral stability in the engineering of hBN emitters \cite{fournier2020position}. Finally, it is worth mentioning that the best quality hBN emitters are still those found randomly in the material \cite{grosso2017tunable,dietrich2020solid}. At the moment, the site-specific engineered emitters do not demonstrate a comparable linewidth or purity on par with identified random defects. These favor the hypothesis that most of the engineering methods are activating certain intrinsic defects rather than creating an isolated defect that is responsible for single-photon emission \cite{aharonovich2016solid}.

\subsection{Electrical control of SPEs}
Electrically-driven single-photon emission promises better integration and more practical deployment for SPEs in photonic and optoelectronic platforms. TMDs are amongst the few hosts for quantum emitters that can benefit from the electrical injection of SPEs forming a quantum light-emitting diode (LED) \cite{palacios2016atomically, schuler2020electrically}. Moreover, electrostatic gating of emitters in TMDs has been useful in externally controlling their emission characteristics \cite{chakraborty2015voltage, hotgergate2021gate}.

\begin{figure*}[t] \centering
     \includegraphics[scale= 0.5]{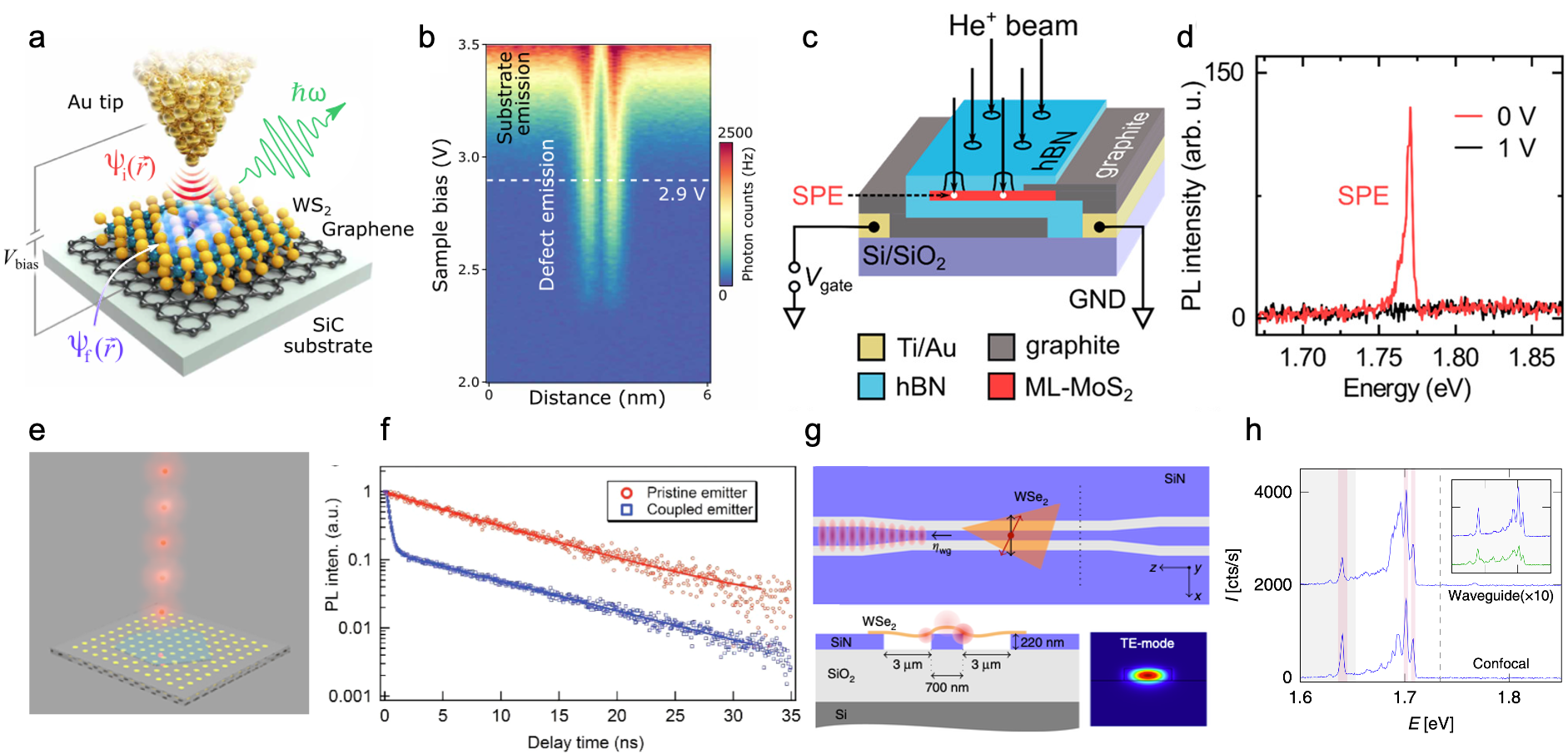} 
          \caption{Electrical control and photonic integration of SPEs. a) A schematic of electrical excitation of SPE by electron tunneling from a gold tip to select defect points \cite{schuler2020electrically}. b) Photon emission across a single defect as function of tunneling bias. c) Gate switching of SPEs in MoS$_2$ monolayer \cite{hotgergate2021gate}. d) Photoluminescence of a selected SPE at gate voltages of 0 V (red curve) and 1 V (black curve). e) An hBN sample on a plasmonic lattice \cite{tran2017deterministic}. f) Time-resolved PL from an uncoupled  emitter (red curve) and an emitter coupled to a plasmonic lattice (blue curve). g) Integrated SPEs in a WSe$_2$ flake with SiN waveguide   \cite{peyskens2019integration}. h) Emission from a confocal and a waveguide-coupled SPEs. The spectrum of the  waveguide-coupled SPEs is multiplied by 10 and offset by 2000 cts/sec for visualization. 
          Panels (a,b): Adapted with permission from \cite{schuler2020electrically}. Copyright 2020, The Authors. Distributed under CC BY-NC. Panels (c,d): Reprinted with permission from  \cite{hotgergate2021gate}. Copyright 2021, American Chemical Society. Panels(e, f): Reprinted with permission from \cite{tran2017deterministic}. Copyright 2017, American Chemical Society. Panels (g,h): Reprinted with permission from \cite{peyskens2019integration}. Copyright 2019, The Author(s). Distributed under the Creative Commons CC BY.
}   
          \label{fig:SPEs_advances}
\end{figure*}

Electrically-excited SPEs have been demonstrated in various vertical and lateral van der Waals heterostructures showing spectrally narrow defect electroluminescence lines similar to that observed under optical excitation \cite{palacios2016atomically, clark2016single, schwarz2016electrically}. Such structures provide an excellent pathway for scalable on-chip quantum technologies; however, they generally lack spatial precision down to the single atomic defect level. Other schemes have been suggested to probe the single-photon emission from individual atomic defects in TMDs using, for example, local electron injection through a gold-coated plasmonic tip \cite{schuler2020electrically} (as shown in Fig. \ref{fig:SPEs_advances}(a)). An example of an optical emission spatial image across a single defect in WS$_2$ as a function of tunneling bias is shown in Fig. \ref{fig:SPEs_advances}(b). Even though the scalability of such a scheme could challenging, it offers a method to correlate the electronic structure of defects with their optical properties using atomic spatial resolution. These measurements would provide insight into the nature and characteristics of the emitters through local electrical perturbations.

The Stark effect has been exploited to tune the optical and electronic properties of defects in TMDs and hBN samples by applying external electric field \cite{chakraborty2017quantum, noh2018stark, xia2019room}, similar to what is routinely implemented with self-assembled quantum dots \cite{patel2010two,somaschi2016near}. The Stark shift is useful in modifying the spectral alignment of the different defects, stabilizing the charge-noise environment of the quantum dot, and also providing valuable information on the dipole moment of the associated defects. Moreover, gate-switching has been used to electrostatically tune the intensity of the photoluminescence of engineered SPEs in MoS$_2$ (Fig. \ref{fig:SPEs_advances}(c)), effectively leading to turning on and off the quantum emission from the emitters as can be observed in Fig. \ref{fig:SPEs_advances}(d) \cite{hotgergate2021gate}. The quantum emitters are sensitive to the charge carrier concentration, which affects the Coulomb interactions on the SPEs. This screening of the localized exciton is responsible for the switching of  single-photon emission between the on and off states. 

It is also worth mentioning that some of the emitters in TMDs have anomalously large values of \textit{g}-factors of up to 12 \cite{koperski2015single}. In the Faraday configuration where the magnetic field is perpendicular to the sample, the emitter's energy can be tuned across several hundreds of $\mu$eV using the Zeeman shift\cite{clark2016single, koperski2015single, chakraborty2015voltage}. Other emitters show no significant Zeeman splitting with negligible \textit{g}-factors indicating that native emitters in TMDs could originate from different electronic transitions \cite{dang2020identifying}.  
 
\subsection{Integration of SPEs with photonic devices}
In addition to the reliable and high-yield creation of SPEs, the extraction of photons and their coupling to well-defined electromagnetic modes are essential for scalable device technologies. Interfacing SPEs with micro- and nano-photonic cavities can increase the rate of their spontaneous emission through the Purcell enhancement. Coupling to resonant cavities can also increase the indistinguishability of the emitted photons through shortening their lifetime to overcome short-time scale dephasing processes \cite{ding2016demand,liu2018high}. 

Plasmonic nanocavity arrays have been a popular choice for providing Purcell enhancement to SPEs in 2D materials \cite{tran2017deterministic, luo2018deterministic, proscia2019coupling}.  Plasmonic arrays of gold and silver nanoparticles supporting lattice plasmon resonances have been used to enhance the emission of hBN emitters by a factor of 2. The use of plasmonic nanoparticles also shortens the lifetime of the emitter by a factor of 30; see Fig. \ref{fig:SPEs_advances}(e,f). This, however, corresponds to an increase in the saturated count rate only by a factor of 2.6 due to the spatial as well as spectral misalignment of the emitters with respect to the particles \cite{tran2017deterministic}. Similarly, arrays of plasmonic nanocubes supporting gap plasmon have been coupled with $\text{WSe}_2$ emitters leading to an average lifetime reduction by a factor of 15 and an average Purcell factor of 181 with factors up to 551 \cite{luo2018deterministic}. The edges of the plasmonic nanocubes also induce strain that is leveraged to localize the SPEs with a high success rate of around 95\%. This results in at least one SPE per nanocube and an automatic deterministic emitter–mode coupling. Plasmonic structures are effective in providing high field enhancement with sub-wavelength confinement. This, however, is generally associated with material losses from the plasmonic components, which prompts the search of an  all-dielectric alternative.  

Other cavity designs have also been examined for coupling to 2D SPEs. A dielectric confocal microcavity has proved useful in improving the purity and indistinguishability of hBN SPEs, reducing the spectral width from 5.76 nm to 0.224 nm \cite{vogl2019compact}. Similarly, silicon nitride dielectric photonic crystal cavities have been coupled to hBN SPEs, providing pathways for better scalability through integrating the hBN prior to the fabrication process \cite{froch2020coupling} and tunability that allow precise emitter-cavity coupling \cite{kim2018photonic}.
 
Coupling SPEs to waveguides has also been investigated to realize on-chip light routing with silicon nitride \cite{peyskens2019integration}, lithium niobate \cite{white2019atomically}, and aluminum nitride \cite{kim2019integrated} photonics. Figure \ref{fig:SPEs_advances}(g) shows an example of a monolayer WSe$_2$ coupled to a silicon nitride waveguide. The emission of an SPE from a confocal PL scan versus the PL of a waveguide-coupled emitter is shown in \ref{fig:SPEs_advances}(h). 
While single-photon emission has been successfully coupled to waveguides in these demonstrations, the coupling efficiency to waveguides remains relatively low. Further optimization of the defect alignment and the field overlap of the single-photon with that of the waveguide is required to enhance the light routing. Moreover, moving beyond proof-of-concept, additional components and functionalities will need to be integrated on-chip to enable efficient quantum information processing and detection. 

\section{\label{sec:prospects}Prospects and challenges}
    \noindent\textbf{Origins of the defects}
    As discussed previously, the deterministic engineering of the spectral properties and the emitters' spatial locations is of paramount importance to their scalability and integration. A key element is understanding the precise origins of the quantum defects and their crystallographic and symmetry properties, which are still not well-understood. The complementary \textit{ab-initio} modeling \cite{ivady2020ab}, combined with Optically Detected Magnetic Resonance (ODMR) and Electron Paramagnetic Resonance (EPR)\cite{gottscholl2020initialization} spectroscopy, have elucidated the nature of the near-infrared hBN emitter to a great extent. However, such studies are lacking for the TMD systems. This is partly because TMD SPEs microscopic origin underlies  more complex interactions of excitons, defects, and strain. Moreover, as suggested by recent experiments\cite{parto2020irradiation, klein2019site}, defects with characteristics potentially distinct from naturally occurring defects can be intentionally created, giving rise to many questions on the extent of control that can be exercised on the properties of engineered quantum emitters in 2D materials. This is not only important for engineering better SPEs, but also can open up the field to other explorations. For instance, can defects with possible high-spin ground-states and spin-photon interfaces be engineered in these materials? TMDs already inherit  dangling bond-free interfaces that allow defect operation at the surface with long spin coherence time \cite{ye2019spin}, strong coupling, and large photon extraction efficiencies that are critical metrics to spin-qubit and quantum sensing applications. On-going efforts on experiments with electron and ion beam irradiation of both hBN and TMDs, combined with first-principles calculation, can help elucidate, assess, and provide guidelines on the various possibilities for engineering extrinsic defects with possible distinct optical properties in 2D materials. 
    

    \noindent\textbf{Indistinguishability}
    Indistinguishable SPEs have not been demonstrated in any 2D materials to date. Indistinguishability (0 <$\xi$< 1) at short time scales can be measured by quantifying the extent to which an emitter's linewidth is lifetime limited,  $ \xi = (T_1  \Gamma )^{-1} $ where $T_1$ and $\Gamma$ denote the emitters lifetime and FWHM line-width, respectively. Remarkably, even though lifetime-limited emitters have been demonstrated in hBN at room temperature \cite{dietrich2020solid}, due to the spectral diffusion, the indistinguishably has not been measured yet. The pathway for  indistinguishable WSe$_2$ and other TMD emitters seems promising.  . Linewidth of SPEs in TMDs are now approaching $10 \mu eV $ range in high-quality and engineered TMDs with lifetimes around $\sim 2$ ns (0.32 $\mu$ eV Fourier-limited linewidth). These metrics may improve with resonant excitation techniques. Furthermore, integrating SPEs with well-designed cavities with Purcell enhancement can further reduce the lifetime ($T_1$) of emitters such that the linewidth becomes lifetime limited, improving the indistinguishability.
    
    
\begin{figure*}[hbt!] 
     \includegraphics[scale= 0.45]{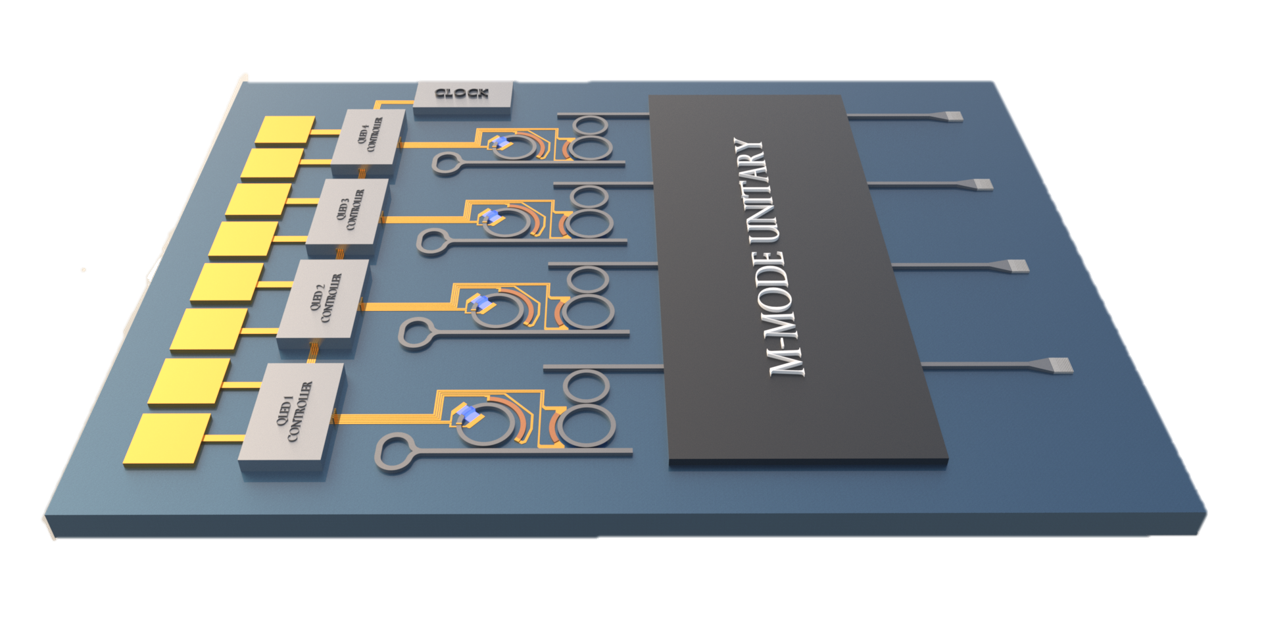} 
          \caption{Future 2D Quantum Light Sources. An illustration for future  on-chip 2D quantum light sources. 2D materials shown in blue can be engineered to become electrically addressable (Quantum LED) and simultaneously integrated with tunable micro-ring resonators. Followed by a tunable Vernier ring filter to ensure isolation of the single-photon-emitter. The array of SPEs are timed by on-chip electronics and can be fed into an arbitary quantum photonic chip for various quantum information applications. Figure is not drawn to scale.  }   
          \label{fig:future_2d}
\end{figure*}

     \noindent\textbf{Photonic interfaces}
     Integrating SPEs with photonic structures is crucial for scalable, stable, and on-chip performance. Interfacing SPEs in 2D hosts with photonic and optoelectronic devices can drastically enhance their performance and open up opportunities to engineer their properties alongside their interaction with the environment. Recent advances in novel cavity designs with unique characteristics have generated significant interest leading to unprecedented functionalities such as polarized cavities \cite{wang2019towards}, chiral cavities \cite{hubener2020engineering}, and topological cavities \cite{gao2020dirac}. The electromagnetic fields in these cavities are associated  with unique properties that can ultimately be imprinted on the hybrid atom-photon state. SPEs in 2D materials are well-poised to benefit from these advancements due to the straightforward integration using dry-transfer techniques and the accessibility of the emitters. Advanced cavity designs can enhance the emission rate of the SPEs, manipulate the emission's polarization, enhance their nonlinear interactions, and even realize non-equilibrium states of matter.

    Moreover, coupling the emission of SPEs to waveguides is also essential for on-chip routing of single photons produced on-demand. Even though vertical integration of 2D heterostructure on the waveguide is straightforward, the electromagnetic mode confinement inside of the waveguide makes it challenging to achieve mode overlap with the emitters. Advanced mode engineering is required to realize significant field overlap and appreciable coupling between the single-photon emission and the guided modes in photonic waveguides. This could be achieved by exploring  waveguide geometries tailored for better field coupling as well as optimizing the location and orientation of the dipole emitter.  Similarly, aligning the polarization of the emitter with that of the waveguide or the cavity is vital for maximum coupling efficiency. The anisotropic deformation induced by the strain can cause rotation of the emitter's dipole moment leading to efficient rotation of its polarization \cite{so2021polarization}. A deeper understanding of the effect of the different types of the strain and how it can be used to control the dipole moment alignment of the emitter would be beneficial for on-chip operation.

    \noindent\textbf{Room-temperature functionality} 
    An emitter's ability to function at room temperature can be a significant advantage when comparing the practicality of various quantum light technologies. In 2D materials, hBN emitters are functional at room temperature due to the large bandgap of hBN, partially isolating the internal transitions of the defects from the environment. On the other hand, initial demonstrations of WSe$_2$ emitters were limited to cryogenic temperatures, and emission would quench at temperatures above 30 K. This quenching behavior, reminiscent of  self-assembled III-V quantum dots\cite{le2003temperature}, can be attributed to the emitters' low confinement potential that makes the quantized states prone to depopulation via coupling to phonons. At the simplest level, one can fit the quenching of the PL within an Arrhenius model where $I(t) = I_0 / (T_r/T_{nr}+exp(-E_a/kT))$, where $T_r$, $T_{nr}$, and $E_a$ are the radiative recombination lifetime, non-radiative recombination lifetime, and energy barrier, respectively. Therefore, it can be seen that the emitters working temperature can be improved by three main strategies, a) increasing the confinement potential (high $E_a$), b) minimizing the non-radiative pathways (high-quality materials), and c) by increasing the radiative recombination rate through Purcell enhancement (reduction of emitter radiative lifetime). Liu \textit{et al}. \cite{luo2019single} successfully increased the working temperatures of WSe$_2$ emitters to 160 K by incorporating both Purcell enhancement and high-quality WSe$_2$. On the other hand, e-beam induced defects in WSe$_2$ that appear deeper in the bandgap can potentially retain their quantum nature up to 215 K \cite{parto2020irradiation}. Overall, these results show that careful engineering of WSe$_2$ emitters combined with well-designed cavities can potentially enable room-temperature functionality in the near future.
    
   \noindent\textbf{Electrically addressability} Recent demonstrations of SPE LEDs in WSe$_2$ show a promising pathway forward \cite{palacios2016atomically,clark2016single,schwarz2016electrically}, especially given the advancement in the site-specific placement of emitters in TMDs; mature TMD single-photon LEDs with minimized background emission are expected in the short term. Here, the challenges compound for hBN: given its insulator nature, creating a \textit{p-n} junction in the material is challenging. Furthermore, as discussed, SPEs in hBN originate from internal defect/dangling-bond transitions. While LEDs based on defect transitions have been previously demonstrated in diamond \cite{lohrmann2011diamond,mizuochi2012electrically} and silicon carbide \cite{lohrmann2015single}, some fundamental questions regarding electrically driving a defect transition still withstand. For instance, how does the variation of the charging state avoided in an electrically active \textit{p-n} junction? Finally, while electrical excitation is encouraging from a scalability point of view, the single-photon purities of electrically excited SPEs are not yet comparable to photoexcitation schemes. This is most evident compared to resonantly photoexcited emitters, which can reduce the background emission and minimize spectral wandering and pure dephasing. Approaches to resonant electrical excitation have been proposed to mitigate this gap, but they remain unexplored in 2D SPEs. Considering that 2D materials family has become a flexible test-bed for proof-of-concept electro-optical devices, it might be timely to investigate novel 2D device structures capable of resonant electrical excitation.

    
    \noindent\textbf{Entanglement}
    Entangled light sources have not been demonstrated in any 2D material yet. One method for entangled-photon emission, using self-assembled quantum dots, for example, is through excitation of a biexciton complex, which is then subsequently recombined, generating two polarization-correlated single photons. Biexciton complexes have been detected in 2D TMDs \cite{you2015observation,he2016cascaded}. However, all of the observations feature a non-zero fine-structure splitting of the intermediate exciton transition. While this provides "which path" information for polarization entanglement, one can instead utilize the biexciton-exciton radiative cascade with pulsed excitation for time-bin entangled photon pair generation; however, it is only in the absence of spin splitting that the radiative cascade from the biexciton will act as a polarization-entangled photon pair with high fidelity. Application of strain and electric field has shown promise to tune the spin splitting in quantum dots but are yet to be investigated in 2D materials. Finally, as we noted before, in TMDs, it is possible that several types of defects with unique defect-bound excitons' optical properties can serve as quantum emitters. Therefore, it might be possible to find defects with morphologies that give rise to biexcitonic features with minimal fine-structure splitting. 


\section{\label{sec:Conclusion}Conclusion}
The first SPEs in 2D materials were reported six years ago, and ever since, the amount of progress in the field has been remarkable. Considering that high-quality single-photon emission from III-V quantum dots was only achieved after nearly three decades, the rapid trajectory of ever-improving SPEs in 2D materials is promising. SPE metrics such as brightness, purity, and indistinguishability have been improved by orders of magnitude since their discovery. Given the pace of these advancements, 2D SPEs can potentially become a technologically relevant candidate for solid-state quantum light sources in the future. In this perspective, we aimed to emphasize the necessary steps required to engineer and integrate these emitters with on-chip photonic technologies. The advancement in 2D material growth, defect characterization, and \textit{ab-initio} theoretical investigations will continue to improve the emitter's intrinsic metrics in the long term. In the coming years, exploring compatible designs and integration with micro-cavities is necessary to improve the stability, brightness, and indistinguishability of the emitters. Special focus should be directed to  compatible designs that can combine and leverage most of the unique advantages of 2D materials. One such design and our vision for 2D quantum light sources' future is illustrated in Fig.5. WSe$_2$ based LEDs integrated with a tunable micro-ring resonator for Purcell enhancement and tunable Vernier spectral filters can provide a scalable pathway for on-chip quantum light sources.  It is readily observable that such integration requires a thorough understanding of site-specific engineering of WSe$_2$ emitters with high SPE qualities, integration with compatible cavity designs with appropriate Purcell enhancement, and engineering of LEDs capable of resonant electrical excitation. Prioritized research and development in each of these areas may accelerate the pace towards arrays of on-demand, room-temperature sources of indistinguishable on-chip single-photons.\\

\noindent \textbf{Acknowledgements} We gratefully acknowledge support from the UC Santa Barbara NSF Quantum Foundry funded via the Q-AMASE-i program under award DMR-1906325. S.I.A acknowledges support from the California NanoSystems Institute through the Elings fellowship. S.I.A. and G.M. also acknowledge support from NSF ECCS-2032272.\\

\noindent \textbf{Data Availability} Data sharing is not applicable to this article as no new data were created or analyzed in this
study.

\bibliography{references}

\end{document}